\documentclass[12pt,onecolumn]{IEEEtran}
 \IEEEoverridecommandlockouts

\usepackage{arydshln}
\usepackage{float}
\usepackage[colorlinks,linkcolor=blue]{hyperref}
\usepackage[latin1]{inputenc}
\usepackage{amsmath}
\usepackage{graphicx}
\usepackage{amssymb}
\usepackage{amsthm}
\usepackage{verbatim}
\usepackage{epsfig}
\usepackage[labelfont=bf]{caption}
\usepackage{enumerate}
\usepackage{subfig}
\usepackage{epstopdf}

\usepackage{booktabs} 
\usepackage{algorithmicx,algorithm}

\newtheorem{definition}{Definition}
\newtheorem{assumption}{Assumption}
\newtheorem{lemma}{Lemma}
\newtheorem{remark}{Remark}

\newtheorem{proposition}{Proposition}

\def\begquo{\begin{quote}}
\def\endquo{\end{quote}}
\def\begequarr{\begin{eqnarray}}
\def\endequarr{\end{eqnarray}}

\def\begarr{\begin{array}}
\def\endarr{\end{array}}
\def\begequ{\begin{equation}}
\def\endequ{\end{equation}}
\def\lab{\label}
\def\begdes{\begin{description}}
\def\enddes{\end{description}}
\def\begenu{\begin{enumerate}}
\def\begite{\begin{itemize}}
\def\endite{\end{itemize}}
\def\endenu{\end{enumerate}}

\def\lef[{\left[\begin{array}}
\def\rig]{\end{array}\right]}
\def\qed{\hfill$\Box \Box \Box$}
\def\begcen{\begin{center}}
\def\endcen{\end{center}}
\def\begrem{\begin{remark}\rm}
\def\endrem{\end{remark}}
\def\begdef{\begin{definition}}
\def\enddef{\end{definition}}
\def\begpro{\begin{proposition}}
\def\endpro{\end{proposition}}
\def\begfac{\begin{fact}}
\def\endfac{\end{fact}}
\def\begass{\begin{assumption}}
\def\endass{\end{assumption}}
\def\begsubequ{\begin{subequations}}
\def\endsubequ{\end{subequations}}

\def\begmat#1{\begin{bmatrix}#1\end{bmatrix}}
\def\begali#1{\begin{align}{#1}\end{align}}
\def\begalis#1{\begin{align*}{#1}\end{align*}}

\def\caly{{\cal Y}}

\def\L2e{{\cal L}_{2e}}

\def\rea{\mathbb{R}}

\def\sign{\mbox{sign}}

\def\adj{\mbox{adj}}
\def\col{\mbox{col}}

\def\ARC{{\it Annual Reviews in Control}}

\def\TAC{{\it IEEE Trans. Automatic Control}}

\def\AUT{{\it Automatica}}

\usepackage{color}

\usepackage[prependcaption,colorinlistoftodos]{todonotes}

\input epsf

\title{Distributed Observers for LTI Systems with Finite Convergence Time: A Parameter Estimation-based Approach}

\author{Romeo Ortega \quad Emmanuel Nu\~no\quad Alexei Bobtsov 
\thanks{R. Ortega is with the Universit\'e Paris-Saclay, CNRS, CentraleSup\'elec, Laboratoire des Signaux et Syst\`emes, 91190, Gif-sur-Yvette, France and the Department of Control Systems and Robotics, ITMO University, Kronverkskiy av. 49, Saint Petersburg, 197101, Russia. E. Nu\~no is with the Dept. of Computer Science CUCEI, at University of Guadalajara, Guadalajara, Mexico. A. Bobtsov is with the Department of Control Systems and Robotics, ITMO University, Kronverkskiy av. 49, Saint Petersburg, 197101, Russia.  E-mails: ortega@lss.supelec.fr; emmanuel.nuno@cucei.udg.mx; bobtsov@mail.ru}
}

\begin{document}

\maketitle

\begin{abstract}
 	A novel approach to solve the problem of distributed state estimation of linear time-invariant systems is proposed in this paper. It relies on the application of parameter estimation-based observers, where the state observation task is reformulated as a parameter estimation problem. In contrast with existing results our solution achieves convergence in finite-time, without injection of high gain, and imposes very weak assumptions on the communication graph---namely the existence of a Hamiltonian walk. The scheme is shown to be robust {\em vis-\'a-vis} external disturbances and communication delays. 
\end{abstract}



%
\section{Introduction}
\lab{sec1}
%
In this paper we consider the problem of distributed state estimation (DSE) where we are given a dynamical system, a number of sensors with computing capabilities measuring part of the systems state, and a communication network connecting the sensors, viewed as nodes of a {\em directed graph} ${\mathcal G}$. The DSE problem is to reconstruct the global state of the system  without the need for a central coordination unit. We consider two variations of this objective, when the full state is to be reconstructed at {\em a distinguished} node and when this happens at {\em every} node---a property called {\em state-omniscience} in \cite{PARMAR}.  DSE has been considered in a wide range of applications, from network localization to environmental monitoring, surveillance, object tracking, collaborative information processing, and traffic monitoring. Even though there are practical applications for both state reconstruction scenarios discussed above we do not dwell on this issue here and only identify the {\em communication patterns} required by each one of them. See \cite{REGetal} for an excellent, comprehensive, and highly didactical, recent tutorial on the topic and \cite{AKYetal} for a survey on sensor networks.

Although the DRE problem has been under intense study for a number of years, only quite recently have appeared provably correct Luenberger observer-based solutions to it---under reasonably non-restrictive assumptions---see \cite{HANetal,KIMetal,MITSUN,PARMAR,REGetal,WANetal} and references therein. The typical scenario for DRE assumes the system to be linear time-invariant (LTI), whose dynamics is described by $\dot x = Ax$, and each agent senses a signal $y^{(i)} = C_i x,\;i=1,\dots,N$, with $N$ the number of agents. To make the observation problem feasible it is assumed that the system is jointly detectable, {\em i.e.}, that the pair $(C,A)$ is detectable, but none of the pairs $(C_i,A)$ verifies this condition. It is typically assumed that  ${\mathcal G}$ is {\em strongly connected}, though results with weaker assumptions, for instance, that every source component is collectively detectable, have been reported \cite{MITSUN,PARMAR}.

The DSE presented in this paper differs from previous works in three respects.
\begenu[{\bf R1.}]
\item The DSEs mentioned above aim at {\em asymptotic} convergence of the state estimation errors. In contrast with this, in the present work we design {\em finite convergence time} (FCT) observers.
\item Our DSE is based on the recently introduced parameter estimation-based observer (PEBO) technique first reported in \cite{ORTetalscl}, and later generalized in \cite{BOBetal}. The main feature of PEBO, and GPEBO, is that the state estimation problem is translated into a problem of {\em parameter estimation}, yielding stronger convergence properties under weaker excitation assumptions---for instance, the FCT feature mentioned above.
\item The vast majority of the reported results aim at the aforementioned omniscience objective, which is achieved fussing the independent estimates from the various nodes via a consensus law. Although it is possible to add this feature to the independent state estimators proposed here, we avoid this solution and concentrate, instead, in the reconstruction of the full state via direct information exchange between the nodes.   
\endenu

Luenberger observer-based DSEs that enjoy the FCT property have already been reported in  \cite{SILetaltac,SILetalecc,SILetalscl}. Our GPEBO-based DSE outperforms these designs due to the following considerations.
\begenu[{\bf C1.}]
\item To achieve the FCT objective all the former DSEs inject---via the use of fractional power error terms---high-gain into the observer. As is well-known high-gain based feedback designs are sensitive to noise. 
\item The DSE of  \cite{SILetaltac} is restricted to the very particular case of multiple chains of integrators \cite[eqs (12) and (13)]{SILetaltac}. As recognized by the authors this is not a canonical form for the DSE problem---therefore, it is unclear to which classes of systems it is applicable. For this design the authors prove input-to-state-stability with respect to system and measurement noise, a property that does not rule out the performance degradation due to the presence of noise. 
\item A critical technical assumption on stabilizability of the overall DSE is made in \cite[Assumption 1]{SILetaltac}. This assumption involves the system matrices as well as the graph topology in a far from clear form. 
\item In \cite{SILetalecc} the aforementioned structural assumption of the system is relaxed, but an unusual condition---namely, that the unobservable states of each node can be expressed as a linear combination of the observable states of its neighboring nodes \cite[Assumption 1]{SILetalecc}---is imposed. As recognized by the authors the validity of this assumption depends on the graph topology, and even on the output matrices of each node. It is unclear under which conditions on the graph this assumption is satisfied, besides the trivial case of a {\em complete} graph. 

\item In \cite{SILetalscl} an individual observability decomposition similar to the one of \cite{KIMetal} is used and---likewise \cite{HANetal,PARMAR,WANetal}---the convergence proof requires a {\em strongly connected} graph.  
\endenu  

In our DSE design there is no injection of high-gain, no particular assumptions---besides observability (or detectability)---on the matrices $(C,A)$ , and the only requirement we impose on the graph is the existence of a {\em Hamiltonian walk}. 

To achieve the FCT property we do not estimate the unknown parameters of the GPEBO with a classical gradient (or least-squares) estimators, that ensure only asymptotic convergence \cite{SASBOD}. Instead we adopt the dynamic regressor extension and mixing (DREM) approach first proposed in \cite{ARAetaltac}.\footnote{See \cite{ORTetalaut} for an interpretation of DREM as a functional Luenberger observer.} The main feature of this approach is that a $q$-dimensional linear regression equation (LRE) is transformed into $q$ {\em scalar} LREs, for which it is possible to ensure the FCT property. Another property of DREM is that FCT is ensured under the extremely weak assumption of {\em sufficient excitation} \cite{KRERIE}---that should be contrasted with the stringent persistency of excitation requirement of standard estimators \cite[Theorem 2.5.1]{SASBOD}. 

Our DSE strongly relies on the interesting observability decomposition introduced in \cite{MITSUN}. This decomposition of observable and unobservable subspaces for each sensor is instrumental to permit each sensor to estimate only its observable sub-space. In \cite{MITSUN} these estimated subspaces are diffused via a consensus law. In contrast with this, in our GPEBO design we use the staircase form of the transformed system to transfer, ``from the upward nodes down", this information directly. Thanks to the FCT property of the estimates the downward nodes will receive the correct information  in finite time. A drawback of this construction is that, since convergence occurs sequentially, the performance and convergence rate of the overall DSE depends on the topology of the network and on how the sensors are ordered. This problem does not ocurr in the highly original work of \cite{KIMetal}---or its refined versions \cite{HANetal,WANetal}---where a similar subspace decomposition is used. See also \cite{DELetal} and \cite{SILetalscl}  for an FCT version.   

The remainder of the paper is organized as follows. In Section \ref{sec2} we present the problem formulation and explain the basis of GPEBO. To set-up the notation, in Section \ref{sec3} we introduce the observability decomposition \cite{MITSUN}.  Section \ref{sec4} contains the main result of the paper, namely the design of our FCT DSE. In Section \ref{sec5} we discuss some extensions and wrap-up the paper with simulation results in Section \ref{sec6}. For the sake of completeness we give in the Appendix an alternative solution to the omniscience problem using an FCT consensus algorithm  
%
\section{Problem Formulation and Introduction to GPEBO}
\lab{sec2}
%
\subsection{The DSE problem}
\lab{subsec21}
%
Consider an  LTI system that, for simplicity, we assume is stable and without inputs. That is
\begsubequ
\lab{sys}
\begali{
\lab{dotx}
\dot x&=Ax\\
\lab{y}
y&=Cx,
}
\endsubequ
with $x \in \rea^n$, $y \in \rea^m$. It is assumed that the state of the system is monitored by a network of $N>1$ sensors, each of which receives a partial measurement of the states
\begequ
\lab{yi}
y^{(i)}=C_i x,\;i \in \bar N :=\{1,2,\dots,N\},
\endequ
where\footnote{Throughout the note the index $i$ ranges in the set $\bar N$, therefore this clarification is omitted in the sequel.} $y^{(i)} \in \rea^{m_i}$, $\sum_{i=1}^N m_i=m$ and $C^\top =\begmat{C_1^\top ,\dots,C_N^\top }$. We assume that the pair $(C,A)$ is {\em observable} but none of the pairs $(C_i,A)$ is observable. The sensors are represented as nodes of an underlying directed communication graph ${\mathcal G}$, which governs the information flow between the sensors. 

The task is to design a distributed algorithm that guarantees either one of the following objectives.
\begenu[{\bf O1.}]
\item Reconstruction of the entire state $x$ at a {\em given} node.
\item Omniscience, that is reconstruction of the entire state $x$ at {\em each} node.
\endenu

In both cases, the convergence should be achieved in {\em finite time}. The vast majority of the papers on DSE concentrate on the---aesthetically appealing---omniscience objective {\bf O2}. However, because of the heavy demands on information exchange, this does not seem to be a ``good" solution in practical applications. On the other hand, there are important practical scenarios where the desired objective is {\bf O1}. For instance, monitoring multiple robotic (underwater) vehicles from one central (surface or on-board) node. 

We make the following assumptions about the interconnecting graph.

\begin{assumption}\label{ass1}\em
For the Objective {\bf O1} it suffices that the graph ${\mathcal G}$ contains an {\em open Hamiltonian walk}.  For Objective {\bf O2} we assume the graph ${\mathcal G}$  contains a {\em closed} Hamiltonian walk.
\end{assumption}

We recall that a walk is a sequence of vertices and arcs of a graph, it is Hamiltonian if it contains {\em every} vertex of the graph \cite[Chapter 1.4]{BANGUTbook}. The walk is closed if the first vertex is equal to the last one, otherwise it is open. The definition of a walk does not rule out the possibility of visiting a node more than one time, which might be impractical. To avoid this situation we need to replace Assumption 1 by having a Hamiltonian (open or close) {\em path}, where all vertices of the walk are {\em distinct}. 

We make the observation that the standard assumption of strongly connected graph made in the DSE papers \cite{HANetal,KIMetal,MITSUN,PARMAR,REGetal,WANetal} is {\em strictly stronger} than the conditions of Assumption \ref{ass1}, since the former {\em implies} the latter but the opposite is not true. It is also significantly weaker than the assumption of {\em proper or complete} graph implicitely made in  \cite{SILetaltac,SILetalecc}.
\subsection{The PEBO approach}
\lab{subsec22}
%
As indicated in the Introduction we will follow the PEBO approach \cite{ORTetalscl} where we translate the state observation problem into a problem of {\em parameter} estimation. In the GPEBO approach \cite{BOBetal}, the  key step is as follows. As is well known \cite[Property 4.1]{RUGbook}, the {\em state transition matrix} of $A$ is the unique, full rank $(n \times n)$ matrix solution of the equation
$$
\dot \Phi =A\Phi,\;\Phi(0)=I_n
$$
with $I_n$ the $(n \times n)$ identity matrix, and is given by
\begequ
\Phi(t) =e^{At}.
\lab{dotcalm}
\endequ 
It has the property that, for any initial condition $x_0=x(0) \in \rea^n$ of the system \eqref{sys}, its solution satisfies
\begequ
\lab{xcalmthe}
x(t)=e^{At}\theta, 
\endequ
where $\theta:=x_0$. Following the GPEBO approach \cite{BOBetal,ORTetalscl} we will treat $\theta$ as an {\em unknown} constant vector to be estimated, in a distributed way, from the measurement of the signals \eqref{yi}. 
%
\section{The Multisensor Observable Canonical Form \cite{MITSUN}}
\lab{sec3}
%
Instrumental for our developments is the use of the multisensor observable canonical form \cite[eq. (5)]{MITSUN}. To set up the---unavoidably messy---notation we recall briefly here this seminal construction. Towards this end, we first introduce the partitions 
$$
x=\col(x^{(1)}, x^{(2)},\dots, x^{(N)} ),\;y=\col(y^{(1)}, y^{(2)},\dots, x^{(N)} ),
$$
with $x^{(i)} \in \rea^{n_i},\;y^{(i)} \in \rea^{m_i}$, $\Sigma_{i}n_i=n,\;\Sigma_{i}m_i=m$ and the  {\em block triangular} sub-system matrices 
\begali{
\lab{asupi}
A^{(i)}&:= \begmat{A_{11} & 0_{n_1 \times n_2}& \cdots & 0_{n_1 \times n_{(i-1)}} & 0_{n_1 \times n_i}\\ 
                        A_{21} & A_{22} &\cdots & 0_{n_2 \times n_{(i-1)}}  & 0_{n_2 \times n_i}\\ 
                        \vdots & \vdots & \vdots & \vdots & \vdots \\
                        A_{i1} & A_{i2} &\cdots & A_{i(i-1)} &  A_{ii}
                         },
}
from which we get
\begequ
\lab{dotxsupi}
\begmat{\dot x^{(1)}\\ \vdots \\ \dot x^{(i)}}=A^{(i)}\begmat{x^{(1)}\\ \vdots \\ x^{(i)}},
\endequ
and we observe that $A^{(N)}=A$. For the output vectors we define  {block triangular}  output matrices
$$
\begin{aligned}
\begmat{y^{(1)}\\ \vdots \\ y^{(i)}}=&\begmat{C_{11} & 0_{m_1 \times n_2}& \cdots & 0_{m_1 \times n_{(i-1)}} & 0_{m_1 \times n_i}\\ 
                        C_{21} & C_{22} &\cdots & 0_{m_2 \times n_{(i-1)}}  & 0_{m_2 \times n_i}\\ 
                        \vdots & \vdots & \vdots & \vdots & \vdots \\
                         C_{i1} & C_{i2} &\cdots & C_{i(i-1)} &  C_{ii}
                         }
                         \begmat{x^{(1)}\\ \vdots \\ x^{(i)}}\\
                         =: & C^{(i)} \begmat{x^{(1)}\\ \vdots \\ x^{(i)}},
\end{aligned}
$$
again verifying that $C^{(N)}=C$. As shown in \cite{MITSUN}, from the assumption of observability of the pair $(C,A)$ it follows that the pairs $(C_{ii},A_{ii})$ are also {\em observable}. 

As discussed in  Section \ref{sec2}, the first step in the design of the GPEBO for the $i$-th agent, is the computation of the state transition matrix of the associated subsystem \eqref{dotxsupi}. It is easy to see that the matrix $\Phi^{(i)}=e^{A^{(i)}t}$ also has a block upper-triangular form, that is,
\begali{
\lab{phisupi}
\Phi^{(i)}&:= \begmat{\Phi_{11} & 0_{n_1 \times n_2}& \cdots & 0_{n_1 \times n_{(i-1)}} & 0_{n_1 \times n_i}\\ 
                        \Phi_{21} & \Phi_{22} &\cdots & 0_{n_2 \times n_{(i-1)}}  & 0_{n_2 \times n_i}\\ 
                        \vdots & \vdots & \vdots & \vdots & \vdots \\
                        \Phi_{i1} & \Phi_{i2} &\cdots & \Phi_{i(i-1)} &  \Phi_{ii}
                         }.
}
Moreover, we have that 
\begequ
\lab{phiii}
\Phi_{ii}(t)=e^{A_{ii}t}.
\endequ

Notice that, compared with the multisensor observable canonical form of \cite[eq. (5)]{MITSUN}, we have excluded the presence of an {\em unobservable subspace}. This stems from the fact that we have assumed observability of the pair $(C,A)$, instead of the weaker detectability condition. See point {\bf E1} in Subsection \ref{subsec53} for a discussion on this matter.
%
\section{Design of the Distributed FCT-GPEBO}
\lab{sec4}
%
With the agents ordered according to the Hamiltonian walk stated in Assumption \ref{ass1}, we will carry-out the design of the FCT-GPEBO of each agent in the order of the walk.
\subsection{FCT GPEBO for the first agent}
\lab{subsec41}
%
\begin{proposition}\em
\lab{pro1}
Define the FCT GPEBO for the first agent via the DREM parameter estimator
\begsubequ
\lab{gpebo1}
\begali{
\dot Y_1&=-\lambda_1 Y_1 + \lambda_1  \Psi_1^\top y^{(1)},\; Y_{1}(0)=0_{n_1 \times 1} \lab{doty1}\\
\dot \Omega_1&=-\lambda_1 \Omega_1 + \lambda_1 \Psi_1^\top  \Psi_1,\; \Omega_{1}(0)=0_{n_1 \times n_1}   \ \lab{dotome1}\\
\dot\omega_1 &=-\gamma_1\Delta_1^2\omega_1,\; \omega_1(0)=1 \lab{dotw1}\\
\dot {\hat \theta}_1&=\gamma_1\Delta_1(\mathcal{Y}_1-\Delta_1\hat \theta_1),\label{dothatthe1}
}
\endsubequ
with $\lambda_1>0,\; \gamma_1>0$, free tuning parameters and 
\begali{
\nonumber
\Psi_1 & =C_{11} e^{A_{11}t}  \\ 
\nonumber
\mathcal{Y}_1 & =\text{adj}\{\Omega_1\}Y_1 \\
\Delta_1 & =\det\{\Omega_1\}
\lab{del1}
}
where $\adj\{\cdot\}$ is the adjunct  matrix.

The state estimate
\begequ
\lab{hatxfct1}
\hat x^{(1)}=\Psi_1 \hat\theta^{(1)}_{\tt FCT}
\endequ
where 
\begequ
\lab{hatthefct1}
\hat\theta^{(1)}_{\tt FCT}:=\frac{1}{1-\omega_1^c}\left[\hat \theta^{(1)}-\omega_1^c\hat\theta^{(1)}(0)\right]
\endequ
and the function $\omega_1^c$ is defined via the clipping function
\begin{align}\nonumber
\omega_1^c=
\begin{cases}
\omega_1~~~~~~~\text{if}~~\omega_1<1-\mu_1\\
1-\mu_1~~\text{if}~~\omega_1\geq 1-\mu_1,
\end{cases}
\end{align}
with $\mu_1 \in (0,1)$ a designer chosen parameter, ensures the existence of a time $t_1^c>0$ such that
\begequ
\lab{hatequx1}
\hat x^{(1)}(t)=x^{(1)}(t),~~\forall t>t_1^c,
\endequ
with all signals bounded.
\end{proposition}

\begin{proof}
From the fact that $\dot x^{(1)}= A_{11} x^{(1)}$ and \eqref{xcalmthe}, we have that 
\begequ
\lab{xcalmthe1}
x^{(1)}= e^{A_{11}t} \theta^{(1)}, 
\endequ
where $\theta^{(1)}:=x^{(1)}(0)$. Our next task is to estimate---in a distributed manner---the parameter $\theta^{(1)}$ so that we can reconstruct the first component of the state $x$. Towards this end, we use the output measurements \eqref{yi} to generate a {\em linear regressor equation} (LRE) as follows
\begali{
\nonumber
y^{(1)} & =C_{11} x^{(1)} \\
\nonumber
&= C_{11}e^{A_{11}t}  \theta^{(1)} \\
& =\Psi_1\theta^{(1)},
\lab{lre1}
}
where we used \eqref{xcalmthe1} and the definition of $\Psi_1$ in \eqref{del1}. Following the DREM methodology \cite{ARAetaltac}, we proceed from the LRE above, and carry out the next operations
	\begalis{
	\Psi_1^\top y^{(1)} &= \Psi_1^\top  \Psi_1  \theta^{(1)} \qquad \qquad \quad\;(\Leftarrow\; \Psi_1^\top \times \eqref{lre1})\\
		{\lambda_1 \over p + \lambda_1}[\Psi_1^\top y^{(1)}] & = {\lambda_1 \over p + \lambda_1}[\Psi_1^\top  \Psi_1]  \theta^{(1)} \qquad (\Leftarrow\;{\lambda_1 \over p + \lambda_1}[\cdot])\\
		Y_1 &= \Omega_1 \theta^{(1)} \qquad \qquad \;\qquad \;\;(\Leftrightarrow\;\eqref{doty1}, \eqref{dotome1})\\
		\adj\{\Omega_1\} Y_1 &= \adj\{\Omega_1\} \Omega_1 \theta^{(1)} \qquad \;\quad(\Leftarrow\;\adj\{\Omega_1\} \times)\\
		\caly_1 &= \Delta_1 \theta^{(1)}  \qquad \qquad\qquad\quad (\Leftrightarrow\;\eqref{del1}),
	}
where $p:={d \over dt}$ and, to obtain the last identity, we have used the fact that for any (possibly singular) $(n \times n)$ matrix $M$ we have $\adj\{M\} M=\det\{M\}I_n$. Replacing the latter equation in \eqref{dothatthe1} yields the error dynamics
	\begequ
	\lab{parerrequ1}
	\dot {\tilde \theta}^{(1)}=-\gamma_1\Delta_1^2  \tilde \theta^{(1)},
	\endequ
where $ {\tilde \theta}^{(1)}:=\hat \theta^{(1)} - \theta^{(1)}$.  Since $\Delta_1$ is a {\em scalar}, the solution of the latter equation is given by
	\begequ
	\lab{tilthe1}
	\tilde \theta^{(1)}(t)=e^{-\gamma_1 \int_0^t \Delta_1^2(s)ds}\tilde \theta^{(1)}(0),\;\forall t \geq 0.
	\endequ 

We proceed now to prove the FCT property of the DSE. Towards this end, notice that the solution of \eqref{dotw1} is
$$
w_1(t)=e^{-\gamma_1 \int_0^{t}\Delta_1^2(s)ds}.
$$
The key observation is that, using the equation above in \eqref{tilthe1}, and rearranging terms we get that 
\begequ
\lab{keyrel}
[1-w_1(t)]\theta^{(1)}= \hat \theta^{(1)}(t) - w_1(t) \hat \theta^{(1)}(0).
\endequ
We will now prove that there exists a time $t_1^c>0$ such that the {\em sufficient excitation} condition \cite{KRERIE}
\begequ
\lab{sufexc1}
\int^{t_1^c}_0 \Delta_1^2(\tau)d\tau\geq-\frac{1}{\gamma_1}\ln(1-\mu_1),
\endequ
is satisfied. If this is the case,
\begequ
\lab{w1c}
w_1^{c}(t)=w_1(t),\;\forall t > t_1^c.
\endequ
Clearly, \eqref{keyrel} and \eqref{w1c} imply that 
$$
\frac{1}{1-\omega_1^c(t)}\left[\hat \theta^{(1)}(t)-\omega_1^c(t)\hat\theta^{(1)}(0)\right]=\theta^{(1)},\;\forall t > t_1^c,
$$
that, in view of \eqref{hatxfct1}, \eqref{hatthefct1} and \eqref{xcalmthe1}, implies \eqref{hatequx1}.
 
To prove our claim that the {sufficient excitation} condition \eqref{sufexc1} is satisfied, notice that the matrix $\Psi_1^\top  \Psi_1$ is  of the form
\begalis{
\Psi_1^\top  \Psi_1&= e^{A^\top _{11}t}C^\top _{11}C_{11}e^{A_{11}t}.
}
On the other hand, it is known  \cite[Theorem 9.8]{RUGbook} that observability of the pair $(C_{11},A_{11})$ is {\em equivalent} to positivity of the observability Gramian
$$
\int_0^t \Psi_1^\top(s)  \Psi_1(s) ds,
$$
for all $t>0$. The proof is completed noting from \eqref{dotome1} that
$$
\Omega_1(t)=\lambda_1 e^{-\lambda_1 t} \int_0^t  e^{\lambda_1 s}\Psi_1^\top(s)  \Psi_1(s) ds,
$$
which is {\em positive definite} for all $t>0$. Consequently, its determinant is bounded away from zero, {\em i.e.}, $\Delta_1(t) \geq \varepsilon_1 >0$ for all $t> 0$. This ensures that there exists a time $t_1^c$ such that \eqref{sufexc1} holds. 

Boundedness of all signals follows from the fact that $A_{11}$ is a stable matrix---completing the proof.  

\end{proof}

From the definition of the sufficient excitation condition \eqref{sufexc1} we see the role played by the constants $\gamma_1$ and $\mu_1$. Indeed, the right hand side decreases increasing $\gamma_1$ or choosing $\mu_1$ close to one---and this in its turn decreases the convergence time $t_1^c$. However, there are other considerations to take into account in the choice of these numbers. On one hand, $\gamma_1$ is the parameter adaptation gain that, as seen in \eqref{tilthe1}, determines its speed of convergence, so one might be tempted to pick a large number. Nevertheless, from identification and adaptive control theories \cite{SASBOD}, it is well-known  that selecting large values for it brings deleterious effects in the face of noise or unmodelled dynamics. On the other hand, notice that in the time interval $[0,t_1^c]$ the FCT estimated parameter \eqref{hatthefct1} takes the form
$$
\hat\theta^{(1)}_{\tt FCT}:=\frac{1}{\mu_1}\left[\hat \theta^{(1)}-(1 - \mu_1)\hat\theta^{(1)}(0)\right].
$$
Therefore, if we choose the constant $\mu_1$ close to zero there is a potential {\em high-gain} injection. Therefore, there is a compromise in the choice of these constants that---in the absence of clear guidelines---is usually done via trial-and-error. A similar remark should be made concerning the time constant of the filters \eqref{doty1} and \eqref{dotome1}, which is determine by $\lambda_1$. It's choice should be related with the operating bandwidth of the system.    
\subsection{FCT GPEBO for the second agent}
\lab{subsec42}
%
The observer for the second agent is trickier because the LRE associated to $y^{(2)}$ contains also the vector $\theta^{(1)}$. Indeed, from the fact that
$$
\begmat{x^{(1)}\\ x^{(2)}} = \Phi^{(2)} \begmat{ \theta^{(1)} \\\theta^{(2)}},
$$
and proceeding as done above to generate the LRE \eqref{lre1}, we have
\begalis{
y^{(2)}&=C_{21}x^{(1)}+C_{22}x^{(2)}\\
&= (C_{21} \Phi_{11}  + C_{22} \Phi_{21}) \theta^{(1)} +  C_{22} \Phi_{22}\theta^{(2)}.
}
We might be tempted to treat this as an (extended) LRE, that is, writing the latter equation as $y^{(2)} = \bar \Psi_2  \begmat{ \theta^{(1)} \\\theta^{(2)}}$. There are two drawback to this approach. First, that a consistent estimate of  $\theta^{(1)}$ is being already generated by the first agent, so there is no point in estimating it again. Second, the key matrix  $\bar \Psi_2^\top \bar \Psi_2$ takes the form
$$
\bar \Psi_2^\top \bar \Psi_2= \begmat{ \star  & \star  \\ \star  & e^{A^\top _{22}t}C^\top _{22}C_{22}e^{A_{22}t}}.
$$
Similarly to the first node, because of observability of the pair $(C_{22},A_{22})$, upon filtering the matrix $\bar \Psi_2^\top \bar \Psi_2$, its $(2,2)$ block is positive definite. However, there is nothing we can say---{\em a priori}---about the rank of the filtered matrix $\bar \Psi_2^\top \bar \Psi_2$. And it is the determinant of this matrix that determines, via \eqref{tilthe1}, the convergence properties of the parameter estimator. In the extreme, fully decoupled case, when $A_{21}$ and $C_{21}$ are zero, the $\star$ blocks are zero and the filtered matrix is rank deficient.  Of course, the rank may also drop even if $A_{21}$ and $C_{21}$ are nonzero. 

To ensure that we generate a suitable LRE for $\theta^{(2)}$ only we  and define a ``perturbed" LRE as
$$
y^{(2)} - (C_{21} \Phi_{11}  + C_{22} \Phi_{21})\hat\theta^{(1)}_{\tt FCT} = C_{22} \Phi_{22}\theta^{(2)} - (C_{21} \Phi_{11}  + C_{22} \Phi_{21}) \tilde \theta^{(1)}_{\tt FCT}
$$
where $\tilde\theta^{(i)}_{\tt FCT}:=\hat\theta^{(i)}_{\tt FCT} - \theta^{(i)}$, and $\hat\theta^{(1)}_{\tt FCT}$ is information sent from node one to node two. The latter equation can be rewritten as
\begequ
\lab{lre2}
{\mathtt y}^{(2)}={\Psi}_2 {\theta}^{(2)}+\epsilon_2,
\endequ
where 
\begali{
\nonumber
{\mathtt y}^{(2)} &:= y^{(2)} - (C_{21} \Phi_{11}  + C_{22} \Phi_{21})\hat\theta^{(1)}_{\tt FCT}\\
\nonumber
{\Psi}_2&= C_{22} \Phi_{22}\\
\lab{phi2}
\epsilon_2 &:=-(C_{21} \Phi_{11}  + C_{22} \Phi_{21}) \tilde \theta^{(1)}_{\tt FCT}
} 
Notice that, because of the FCT property of the estimate $\hat \theta^{(1)}_{\tt FCT}$, the disturbance term  $\epsilon_2(t)=0,\;\forall t > t_1^c.$ 

We are in position to present the main result for the FCT-GPEBO of the second agent, whose proof follows {\em verbatim} the proof of Proposition \ref{pro1}.\footnote{Since the term $\epsilon_2(t)=0$ for $t > t_1^c$ in the analysis below we set it equal to zero for ease of presentation. See  point {\bf E4} in Subsection \ref{subsec53} for some discussions on the effect of additive vanishing terms in the DREM design.}
 
\begin{proposition}\em
\lab{1pro2}
Define the GPEBO for the second agent via 
\begsubequ
\lab{gpebo2}
\begali{
\dot \Phi^{(2)} &=A^{(2)} \Phi^{(2)},\; \Phi^{(2)}(0)=I_{n_1+n_2} \lab{dotphi2}\\
\dot Y_2&=-\lambda_2 Y_2 + \lambda_2  \Psi_2^\top {\mathtt y}^{(2)},\; Y_{2}(0)=0_{n_2 \times 1} \lab{doty2}\\
\dot \Omega_2&=-\lambda_2 \Omega_2 + \lambda_2 \Psi_2^\top  \Psi_2,\; \Omega_{2}(0)=0_{n_2 \times n_2}   \ \lab{dotome2}\\
\dot\omega_2 &=-\gamma_2\Delta_2^2\omega_2,\; \omega_2(0)=1 \lab{dotw2}\\
\dot {\hat \theta}^{(2)}&=\gamma_2\Delta_2(\mathcal{Y}_2-\Delta_2\hat \theta^{(2)}),\label{dothatthe2}
}
\endsubequ
with $\Psi_2$ and ${\mathtt y}^{(2)}$ defined in \eqref{phi2}, $\gamma_2>0$ and $\lambda_2>0$ free tuning parameters and 
\begali{
\nonumber
\mathcal{Y}_2 & =\text{adj}\{\Omega_2\}Y_2 \\
\Delta_2 & =\det\{\Omega_2\}
\lab{del2}
}

The state estimate
\begequ
\lab{hatxfct}
\hat x^{(2)} =\begmat{\Phi_{21} & \Phi_{22}} \begmat{\hat\theta^{(1)}_{\tt FCT} \\ \hat\theta^{(2)}_{\tt FCT}},
\endequ
where 
$$
\hat\theta^{(2)}_{\tt FCT}:=\frac{1}{1-\omega_2^c}\left[\hat\theta^{(2)}-\omega_2^c \hat\theta^{(2)}(0)\right],
$$
and the function $\omega_2^c$ defined via the clipping function
\begin{align}\nonumber
\omega_2^c=
\begin{cases}
\omega_2~~~~~~~\text{if}~~\omega_2<1-\mu_2\\
1-\mu_2~~\text{if}~~\omega_2\geq 1-\mu_2,
\end{cases}
\end{align}
ensures that
$$
\hat x^{(2)}(t)=x^{(2)}(t),~~\forall t>t_1^c+t_2^c,
$$
for some $t_2^c>0$, with all signals bounded.
\qed
\end{proposition}

\subsection{FCT-GPEBO for the $i$-th agent: $i>2$}
\lab{subsec43}
%
As shown in Subsection \ref{subsec42}, the key step for the design of the  FCT-GPEBO for $i$-th agent, is the generation of the (perturbed) LRE \eqref{lre2} on the $i$-th unknown parameter $\theta^{(i)}$. This result is contained in the lemma below.

\begin{lemma}\em
\lab{lem1}
The (perturbed) LRE of the $i$-th agent is defined as
\begequ
\lab{lrei}
{\mathtt y}^{(i)}={\Psi}_i {\theta}^{(i)}+\epsilon_i,
\endequ
where 
\begali{
\nonumber
{\mathtt y}^{(i)} &:= y^{(i)} - \begmat{ C_{i1}&\cdots & C_{i(i-1)}} \Phi^{(i-1)}\begmat{\hat \theta_{\tt FCT}^{(1)}\\ \vdots \\ \hat \theta_{\tt FCT}^{(i-1)}}\\ 
\nonumber       & - C_{ii}[ \Phi_{i1}\hat \theta_{\tt FCT}^{(1)}+ \cdots+ \Phi_{i(i-1)} \hat \theta_{\tt FCT}^{(i-1)}]  \\
\lab{matttyi}
{\Psi}_i&= C_{ii} \Phi_{ii},
} 
with $\col(\hat\theta^{(1)}_{\tt FCT},\dots,\hat\theta^{(i-1)}_{\tt FCT})$ information sent from the $(i-1)$-th node. Because of the FCT property of these estimates, the disturbance term  $\epsilon_i(t)=0$ in finite time.
\end{lemma}  
\begin{proof}
First, we compute the state transition matrix of the $i$-th subsystem, that is,
$$
\dot \Phi^{(i)}=A^{(i)} \Phi^{(i)}.
$$
Second, we recall that
\begalis{
\begmat{x^{(1)}\\ \vdots \\ x^{(i)}}&= \Phi^{(i)}\begmat{\theta^{(1)}\\ \vdots \\ \theta^{(i)}}\\
&=\begmat{ \Phi^{(i-1)}\begmat{\theta^{(1)}\\ \vdots \\ \theta^{(i-1)}}\\ \Phi_{i1}\theta^{(1)}+ \cdots+ \Phi_{ii} \theta^{(i)}}.
}
Then, we do the following calculations
\begalis{
y^{(i)}&= \begmat{ C_{i1} &\cdots & C_{i(i-1)} &  C_{ii}} \begmat{x^{(1)}\\ \vdots \\ x^{(i)}}\\
       &= \begmat{ C_{i1} &\cdots & C_{i(i-1)} &  C_{ii}} \Phi^{(i)}\begmat{\theta^{(1)}\\ \vdots \\ \theta^{(i)}}\\                 
       &= \begmat{ C_{i1}&\cdots & C_{i(i-1)}} \Phi^{(i-1)}\begmat{\theta^{(1)}\\ \vdots \\ \theta^{(i-1)}}+\\ 
       &+  C_{ii}[ \Phi_{i1}\theta^{(1)}+ \cdots+ \Phi_{i(i-1)} \theta^{(i-1)}]  +  C_{ii}\Phi_{ii} \theta^{(i)}                   
}
As done in Subsection \ref{subsec42}, the proof is completed expressing the parameters as the difference between their FCT estimate and the associated parameter error. 
\end{proof} 

The construction of the $i$-th FCT-GPEBO proceeds from \eqref{lrei} exactly mimicking the construction given in the previous section.

Clearly, the $N$-th node will be able to reconstruct the full state via the estimate
$$
\hat x=\Phi \begmat{\hat \theta_{\tt FCT}^{(1)}\\ \vdots \\ \hat \theta_{\tt FCT}^{(N)}},
$$
which satisfies the FCT objective
$$
\hat x(t)=x(t),\;\forall t > t^c,
$$ 
for some $t^c>\sum_{i=1}^N t^c_i$.
%
\section{Discussion and Extensions}
\lab{sec5}
%
In this section we make some additional remarks about our DSE and discuss some potential extensions of the result given above.
\subsection{Discussion}
\lab{subsec51}
%
\begenu[{\bf D1.}]
\item From the derivations above it is clear that the main idea in our design is that each agent, using it's measurement $y^{(i)}$, generates a LRE for the corresponding parameter $\theta^{(i)}$, which are related with the states $x^{(i)}$. The consistent estimation of these parameters is guaranteed because of the fact that the pairs $(C_{ii},A_{ii})$ are {observable} (or detectable). This is similar to the approach adopted in \cite{MITSUN} where the estimated subspaces are diffused via a consensus law. In contrast with this, in our GPEBO design we use the staircase form of the transformed system to transfer directly this information---encrypted in the parameter estimate $\hat \theta_{\tt FCT}^{(i)}$---from the node $i$ to the node $i+1$ via the definition of ${\mathtt y}^{(i+1)}$ in \eqref{matttyi}. Thanks to the FCT property of the estimates the downward nodes will receive the correct information  in finite time.

\item Another important difference of our approach with respect to  \cite{MITSUN} is the following. In the latter the $i$-th agent is able to estimate {\em only} its associated sub-state, that is $x^{(i)}$. In contrast with this, in our scheme the $i$-th node can reconstruct the sub-vector $\col(x^{(1)},x^{(2)},\dots,x^{(i)})$. Hence, the $N$-th node will estimate the full state $x$. It is clear that, to achieve this objective it is enough that the graph ${\mathcal G}$ has a {\em Hamiltonian walk}, whose terminal edge is the $N$-th node. This assumption is strictly weaker than the standard strict connectivity one. If the Hamiltonian walk is closed, then the information of the full estimated state can be ``transferred back" to the first node and, from there, transmitted to all nodes, achieving the omniscience objective. 

\item To achieve  omniscience, it is also possible to implement in each node $i$ a standard {\em consensus-based dynamics} to estimate the $j$-th substates, for $j \in \bar N \setminus \{i\}$. We notice that, as done in \cite{SILetaltac,SILetalecc}, it is possible to introduce fractional powers in the correction terms to ensure that this consensus term will also converge in finite (or even, fixed) time. In view of the specificity mentioned above---regarding the ``increasing" size of the estimated state---of our construction, it seems that the proposed transfer of the full state estimate to all nodes is more attractive from the computational and reliability view point.  For the sake of completeness, we give in the Appendix a procedure to achieve omniscience using consensus. 

\item It is interesting to observe that there is no need to compute {\em at each node} the corresponding block state transition matrix $\Phi^{(i)}$. It suffices to compute the full matrix, via $\dot \Phi=A \Phi$---or, equivalently,  $e^{At}$---in the first node, and to transfer the blocks of $\Phi^{(i)}$ to the nodes below. 

\endenu

\subsection{Extension to linear time-varying (LTV) systems with inputs}
\lab{subsec52}
%
The GPEBO theory has been developed for nonlinear systems with inputs which, after a change of coordinates, admit an affine representation, see \cite{BOBetal,ORTetalscl}. That is, systems that can be represented by the LTV dynamics
\begalis{
\dot x & = A(t) x + b(t) \\
y&=C(t) x + d(t)
}
where  $A(t),b(t),C(t)$ and $d(t)$ are the evaluation along the trajectories of the input and output signals of the corresponding terms $(\cdot)(t):=(\cdot)(u(t),y(t))$. To apply GPEBO for this variation of the DSE problem  the following modifications to the construction are needed. The first difference is due to the presence of the signals $b(t)$ and $d(t)$. To handle this we define the signal 
\begequ
\lab{exxi0}
e=x-\xi, 
\endequ
with $\xi$ satisfying the equation $\dot \xi  =A(t)\xi+b(t)$. The dynamics of $e$ is described by the LTV system
\begequ
\lab{dote0}
\dot e = A(t)e.
\endequ
Similarly to the LTI case, the columns of the state transition matrix of the system span the space of solutions of \eqref{dote0}. However, the calculation of this matrix, which now has two arguments, is far from obvious. Fortunately, this property is also enjoyed by the  {\em fundamental matrix} of the LTV system \eqref{dote0}, which is the unique solution of the matrix equation
$$
\dot \Phi_A=A(t) \Phi_A,\;\Phi_A(0)=\Phi_A^0 \in \rea^{n \times n},
$$
with $\Phi_A^0$ full-rank, see \cite[Property 4.4]{RUGbook}. More precisely, 
$$
e(t)=\Phi_A(t) [\Phi_A^0]^{-1}e(0),
$$
and we treat $e(0)$ as an unknown parameter $\theta:=e(0)$, that we try to {\em estimate}. Invoking \eqref{exxi0}, the observed state is then generated as
\begequ
\hat x = \xi + \Phi_A \hat \theta,
\endequ
where, to simplify notation and without loss of generality, we set $\Phi_A^0=I_n$.

From the construction of the FCT-GPEBO given in the previous section it is clear that a key component is the availability of the multisensor observable canonical form \cite[eq. (5)]{MITSUN}. Exploiting its structure it was possible to obtain decoupled (perturbed) LREs for the vectors $\theta^{(i)}$, associated to the state of the $i$-th agent, with a ``suitable" regressor matrix $\Psi_i$. It is not clear at this point whether, in the LTV case, it is possible to implement the coordinate transformations that lead to this form. This question is closely related to the availability of coordinate transformations that lead to Kalman's canonical form, see  \cite{JIKHOD} for some recent results on this topic.

\subsection{Robustness properties}
\lab{subsec53}
%
Some {\em robustness} properties of the proposed DSE are easily established. For instance, it is easy to see that, since the regressors $\Delta_i$ are bounded away from zero, the individual agents parameter error dynamics 
$$
\dot {\tilde \theta}^{(i)}=-\gamma_i\Delta_i^2  \tilde \theta^{(i)},
$$
is exponentially stable, hence {\em input-to-state-stable} with respect to external disturbances. 

Also, the presence of transmission {\em delays} (possibly time-varying) between the nodes does not affect the stability of the overall DSE. Indeed, the delays will ``retard" the transfer of ${\hat \theta}_{\tt FCT}^{(i-1)}$ from the $(i-1)$-th node to the $i$-th one, only extending the FCT $t_{i}^c$ of this node. 

Convergence of the overall scheme is ensured even if the graph has a {\em switching topology}, provided the union of the edges of all connectivity graphs contains a Hamiltonian walk for sufficiently long time. However, it should be mentioned that the scheme is highly fragile to the possible loss of one of the agents, but this seems to be the case for all schemes relying on observability decompositions.

\subsection{Further extensions and discussion}
\lab{subsec54}
%
\begenu[{\bf E1.}]
\item We have assumed that the pair $(C,A)$ is observable, in this case there is no unobservability subspace. As done in \cite{HANetal,MITSUN,WANetal} we can relax this assumption to {\em detectability}. In this case, in the multisensor observable canonical form, the lowermost corner of the matrix $C$ is zero and the corresponding term of the $A$ matrix spans the unobservable subspace, see \cite[eq. (5)]{MITSUN}. But, apart from this minor notation modification, the construction of the GPEBO given here, remains unaltered.   

\item Another possibility, alternative to the use of DREM to obtain the scalar regressor $\Delta_i$, is to directly apply a standard gradient estimator to the LRE \eqref{lrei}, that is
$$
\dot {\hat \theta}^{(i)}=\gamma_i\Psi^\top_i(\mathtt{y}_i-\Psi_i\hat \theta^{(i)}),
$$
which, under an assumption of persistency of excitation of the regressor matrix $\Psi_i$, ensures exponential convergence of the observation error \cite[Theorem 2.5.1]{SASBOD}. The interest of introducing DREM is that it enable us to achieve FCT, which is unattainable with a standard gradient estimator. Moreover, this objective is achieved with the assumption of sufficient excitation \eqref{sufexc1}, which is strictly weaker than persistent excitation. 
\item The application of our result to {\em discrete-time} systems follows {\em verbatim} the one given here. The interested reader is referred to \cite{ORTetalifac} where such an algorithm is presented for the more challenging problem of nonlinearly parameterized, separable regressions, that is, when $\mathtt{y} = \Psi \Phi(\theta)$, with $\Phi(\cdot)$ a nonlinear mapping.   
\item The initial conditions of the state transition matrix $\Phi^{(i)}$ and the signals $w_i$ were set to identity. This is done without loss of generality, to avoid the need of dragging the constants $\Phi^{(i)}(0)$ and $w_i(0)$ in the derivations. It is interesting to note that, if we the set the initial conditions of the estimated parameters to zero, the expression of the FCT estimates \eqref{hatthefct1} simplify. Since some prior knowledge might be available on the true value of these parameters, this might affect the performance of the design.   
\endenu
%
\section{Simulation Results}
\lab{sec6}

For the simulations we consider the observed system given by
\begequ
\lab{sysbar}
\dot {\bar x} = \bar A \bar x, \qquad y = \bar C \bar x,
\endequ
where $\bar x\in \rea^6$,
$$
\bar A=\left[\begin{array}{cccccc} 
-1 & 0 & 0 & 0 & 0& 0\\
-1 & 1 & 1 &         0    &     0     &    0\\
    1 &   -2&   -1&   -1&    1 &   1 \\
         0      &   0    &     0  & -1 &         0     &    0\\
   -8 &    1 &   -1 &   -1 &   -2 &         0 \\
    4 &   -0.5 &    0.5&         0 &        0  & -4
\end{array}\right]
$$
and 
$$
\bar C=\left[\begin{array}{cccccc} 
1 & 0 & 0 & 2 & 0& 0\\
2 & 0 & 0 &  1    &     0     &    0 \\\hdashline
2 & 0 & 5 & 0 & 0& 3
\end{array}\right]= \left[\begin{array}{c} \bar C^{(1)}\\\hdashline \bar C^{(2)} \end{array}\right].
$$
These numerical simulations borrow the example of Han et al., \cite{HANetal}, with a slight modification on $\bar C$. Moreover, Han et al. \cite{HANetal} employ four agents for the state estimation, while we only consider two agents, {\em i.e}, $N=2$. Notice that $(\bar C,\bar A)$ is observable, but none of the pairs $(\bar C^{(i)},\bar A)$ is observable.

Now, following the multisensor observable canonical decomposition of \cite{MITSUN} with the change of coordinates $\bar x={\mathcal T}x$ we obtain $A= {\mathcal T}^{-1} \bar A {\mathcal T}$ and $C= \bar C {\mathcal T}$, where
$$
{\mathcal T}=\left[\begin{array}{cccccc} 
-0.7071 & -0.7071 & 0 & 0 & 0& 0\\
0 & 0 & 0 &         1    &     0     &    0\\
0 & 0 & 1 &         0    &     0     &    0\\   
-0.7071 & 0.7071 & 0 & 0 & 0& 0\\
0 & 0 & 0 &         0    &     1     &    0\\
0 & 0 & 0 &         0    &     0     &    1
\end{array}\right],
$$
$$
A=\left[\begin{array}{cc:cccc} 
-1 &        0       &  0    &     0      &   0       &  0\\
         0   &-1    &      0     &    0     &    0      &   0\\\hdashline
    0 &  -1.4142 &   -1 &   -2 &   1 &    1 \\
    0.7071 &    0.7071  &   1 &  1 &       0      &   0\\
    6.3640  &  4.9497 &  -1 &    1 &  -2 &        0\\
   -2.8284  & -2.8284  &   0.5 &  -0.5 &         0 &  -4
\end{array}\right]
$$
and
$$
C=\left[\begin{array}{cc:ccccc} 
-2.1213 & 0.7071 &  0 & 0 & 0& 0\\
-2.1213 & -0.7071 &  0 &  0    &     0     &    0 \\ \hdashline
-1.4142 & -1.4142 &  5 & 0 & 0& 3
\end{array}\right].
$$

Thus, the estimated state is given as $\hat {\bar x}={\mathcal T} \hat x$ and the state of the transformed system and its estimate are divided in two components of size two and four, $x^{(1)}\in\rea^2$, $x^{(2)}\in\rea^4$. $\hat x^{(1)}\in\rea^2$ and $\hat x^{(2)}\in\rea^4$, respectively.

The initial conditions $\bar x(0)=[1 3 -2 -3 -1 2 ]^\top$ are the same as in \cite{HANetal}. The initial conditions of the estimations are ${\hat \theta}_1(0)=0_{2\times 1}$ and ${\hat \theta}_2(0)=0_{4\times 1}$.

The gains and other constants are set as: for agent 1, $\lambda_1=1$, $\gamma_1=5$, $\mu_1=0.05$ and, for agent 2, $\lambda_2=0.8$, $\gamma_2=20$, $\mu_2=0.1$.

\begin{figure}[h]
\begin{center}
\includegraphics[width=0.5\columnwidth]{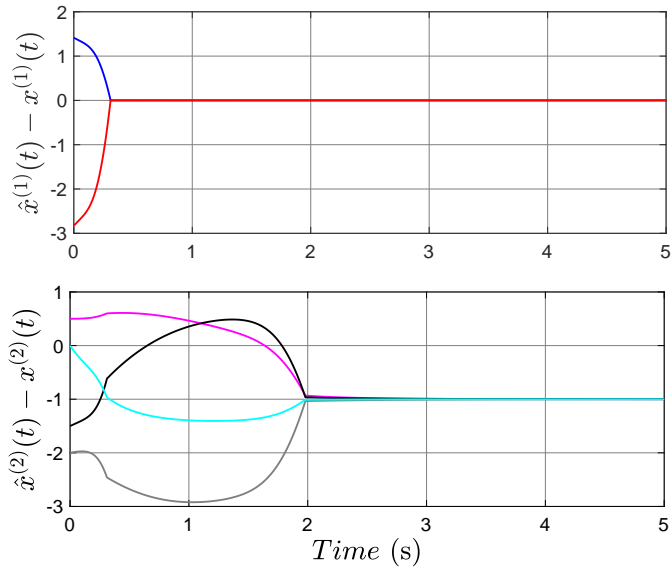}
\caption{Estimation errors of the components of the state $x$.}
\label{error}
\end{center}
\end{figure}

The simulations consider Objective {\bf O1} and the state is reconstructed at Node 2.  Fig.~\ref{error} shows the observation errors for each agent. Recall that Agent 1 estimates two state components and Agent 2 estimates four. As shown in the figure the objective of FCT is achieved with $t_1^c \approx 0.3$s and $t_2^c \approx 1.7$s. Notice that the plots of the estimation error for the second agent experiment a jolt at $t_1 \approx 0.3$s, when the correct estimation of $\theta_{\tt FCT}^{(1)}$ from the first agent happens.

\begin{figure}[h]
\begin{center}
\includegraphics[width=0.5\columnwidth]{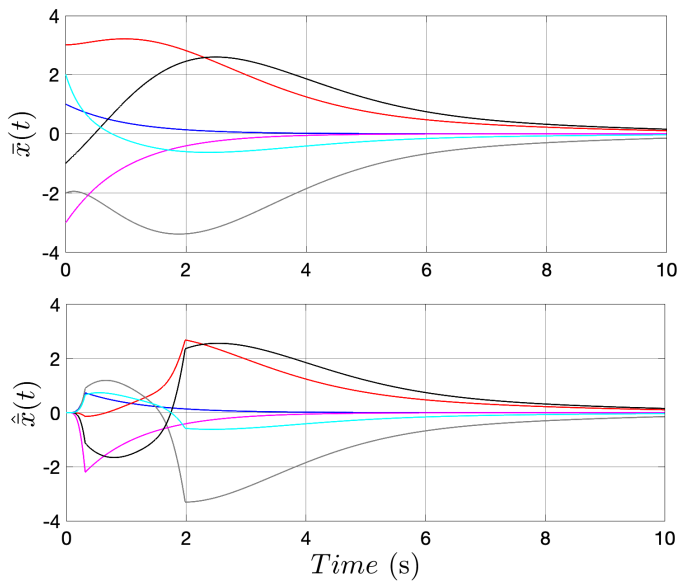}
\caption{Evolution of the state of \eqref{sysbar} $\bar x$ and its observed value $\hat {\bar x}$.}
\label{traj}
\end{center}
\end{figure}

Fig.~\ref{traj} depicts the actual trajectories of the state of the system \eqref{sysbar}, that is, $\bar x$, and its estimates $\hat {\bar x}$. The latter is reconstructed at Agent 2. As predicted by the theory both sets of trajectories coincide after the FCT at $t=2$s.

 %
\appendix
In this appendix we show that, assuming the graph is {\em strongly connected}, we can achieve the omniscience Objective {\bf O2}, implementing a consensus algorithm for each agent $k\in\bar N$ and for each partition of the state $i\in\bar N$. 

Define $z^{(i)}_k\in\rea^{n_i}$ as the corresponding portion of the estimated initial condition $x^{(i)}(0)\in\rea^{n_i}$. Set ${\mathcal N}_k$ as the set of neighbour agents transmitting information to the $k-$th-agent. 
$$
\dot  z^{(i)}_k= -\sum_{j\in{\mathcal N}_k} a_{kj} (z^{(i)}_k - z^{(i)}_k)  -p_k(z^{(i)}_k - \hat\theta^{(i)}_{\tt FCT}),
$$
where $a_{kj}>0$ if $j\in{\mathcal N}_k$ and $a_{kj}=0$, otherwise. Moreover, if $i=k$, then $p_k>0$, otherwise $p_k=0$.

The consensus algorithm can be written also as
$$
\dot  z^{(i)}_k= -\sum_{j\in{\mathcal N}_k} a_{kj} (z^{(i)}_k - z^{(i)}_k)  -p_k( z^{(i)}_k - \theta^{(i)}) - p_k( \theta^{(i)} - \hat\theta^{(i)}_{\tt FCT}),
$$
and defining $z^{(i)}=[(z^{(i)}_1)^\top,\dots, (z^{(i)}_N)^\top]^\top\in\rea^{Nn_i}$ as
\begin{equation}
	\label{dyn_con}
\dot z^{(i)}= - \bar L ( z^{(i)} - 1_N\otimes\theta^{(i)}) + (P\otimes I_{n_i})(1_N\otimes \theta^{(i)} - \hat\theta^{(i)}_{\tt FCT}),	
\end{equation}
where $\bar L:= L + P$, $L\in\rea^{N\otimes N}$ is the standard Laplacian matrix and $P={\rm diag}(p_k)\in\rea^{N\otimes N}$.

Since the graph is strongly connected and using the Gershgorin theorem we can show that all the eigenvalues of $ \bar L $ have strictly positive real parts. Moreover, matrix $ \bar L $ can be always taken to a Jordan normal form. Hence, there exists a proper matrix $\Gamma\in\rea^{N\times N}$ and a block-diagonal matrix $J\in\rea^{N\times N}$ such that
$$ 
 \bar L = \Gamma J \Gamma^{-1}.
$$
Therefore
$$
e^{-\bar Lt} = \Gamma e^{-Jt}\Gamma.
$$
The fact that all the eigenvalues of $ \bar L $ have strictly positive real parts ensures that $e^{-\bar Lt}$ vanishes.

Thus, the consensus dynamics \eqref{dyn_con} can be seen as an exponentially stable linear system perturbed by $1_N\otimes \theta^{(i)} - \hat\theta^{(i)}_{\tt FCT}$. 

However, since $\theta^{(i)} - \hat\theta^{(i)}_{\tt FCT}$ converges to zero in finite-time then $z^{(i)}_k$ converges to $\theta^{(i)}$ exponentially.

The estimation of the state at the $k-$th agent is finally given by
$$
\hat x_k = \Phi z_k,
$$
where $z_k:=[z^{(1)}_k,\dots, z^{(N)}_k]^\top$.
  
Similarly to \cite{SILetaltac,SILetalecc,SILetalscl} consensus can be achieved with FCT using the protocol
$$
\dot  z^{(i)}_k= -\sum_{j\in{\mathcal N}_k} a_{kj} \lfloor z^{(i)}_k - z^{(i)}_k\rceil^r  -p_k\lfloor z^{(i)}_k - \hat\theta^{(i)}_{\tt FCT}\rceil^r,
$$
with $r\in(0,1]$ and $\lfloor a \rceil^r=|a|^r\sign(a)$.

Actually convergence can be achieved in a {\em fix time} using
\begalis{
\dot  z^{(i)}_k &= -\sum_{j\in{\mathcal N}_k} a_{kj} \lfloor z^{(i)}_k - z^{(i)}_k\rceil^{r_1}  -p_k\lfloor z^{(i)}_k - \hat\theta^{(i)}_{\tt FCT}\rceil^{r_1} \\
&-\sum_{j\in{\mathcal N}_k} a_{kj} \lfloor z^{(i)}_k - z^{(i)}_k\rceil^{r_2}  -p_k\lfloor z^{(i)}_k - \hat\theta^{(i)}_{\tt FCT}\rceil^{r_2},
}
with $r_1\in(0,1]$ and $r_2>1$.
%
\subsection*{Acknowledgments}
The first author would like to thank Prof. Nima Monshizadeh for bringing to his attention the concept of Hamiltonian walks and Prof Antonio Pascoal for many insightful remarks about the DSE problem.  This paper is partly supported by the Ministry of Education and Science of Russian Federation (14.Z50.31.0031, goszadanie no. 8.8885.2017/8.9), NSFC (61473183, U1509211); and by the Mexican CONACyT Basic Scientific Research grant CB-282807.
%

\end{document}